\documentclass[draft]{agujournal2019}
\usepackage{url} %this package should fix any errors with URLs in refs.
\usepackage{lineno}
\usepackage{soul}
\usepackage{amssymb,amsfonts,amsmath}
\usepackage{wasysym,gensymb,bm}
\usepackage{mathtools}
\usepackage[normalem]{ulem}
\usepackage{pifont}
\draftfalse

\journalname{Geophysical Research Letters}

\begin{document}

%% ------------------------------------------------------------------------ %%
%  Title
%
% (A title should be specific, informative, and brief. Use
% abbreviations only if they are defined in the abstract. Titles that
% start with general keywords then specific terms are optimized in
% searches)
%
%% ------------------------------------------------------------------------ %%

% Example: \title{This is a test title}

\title{Smaller Sensitivity of Precipitation to Surface Temperature under Massive Atmospheres}

%% ------------------------------------------------------------------------ %%
%
%  AUTHORS AND AFFILIATIONS
%
%% ------------------------------------------------------------------------ %%

% Authors are individuals who have significantly contributed to the
% research and preparation of the article. Group authors are allowed, if
% each author in the group is separately identified in an appendix.)

% List authors by first name or initial followed by last name and
% separated by commas. Use \affil{} to number affiliations, and
% \thanks{} for author notes.
% Additional author notes should be indicated with \thanks{} (for
% example, for current addresses).

\authors{Junyan Xiong\affil{1}, Jun Yang\affil{1}, and Jiachen Liu\affil{1}}

\affiliation{1}{Laboratory for Climate and Ocean-Atmosphere Studies, Department of Atmospheric and Oceanic Sciences, School of Physics, Peking University, Beijing 100871, China}

%% Corresponding Author:
% Corresponding author mailing address and e-mail address:

\correspondingauthor{Jun Yang}{junyang@pku.edu.cn}

%% Keypoints, final entry on title page.

\begin{keypoints}
\item Numerical simulations show that under a given surface temperature, the precipitation is weaker if the air mass is larger.
\item The increasing rate of global-mean precipitation with surface temperature under a larger air mass is also smaller.
\item The combined effect of air mass on Rayleigh scattering, multiple scattering, pressure broadening, and lapse rate is the mechanism.
\end{keypoints}

%% ------------------------------------------------------------------------ %%
%
%  ABSTRACT and PLAIN LANGUAGE SUMMARY
%
% A good Abstract will begin with a short description of the problem
% being addressed, briefly describe the new data or analyses, then
% briefly states the main conclusion(s) and how they are supported and
% uncertainties.

% The Plain Language Summary should be written for a broad audience,
% including journalists and the science-interested public, that will not have
% a background in your field.
%
% A Plain Language Summary is required in GRL, JGR: Planets, JGR: Biogeosciences,
% JGR: Oceans, G-Cubed, Reviews of Geophysics, and JAMES.
% see http://sharingscience.agu.org/creating-plain-language-summary/)
%
%% ------------------------------------------------------------------------ %%

%% \begin{abstract} starts the second page
\justifying
\begin{abstract}
Precipitation and its response to forcings is an important aspect of planetary climate system. In this study, we examine the strength of precipitation in the experiments with different atmospheric masses and their response to surface warming, using three global atmospheric general circulation models (GCMs) and one regional cloud-resolving model (CRM). We find that precipitation is weaker when atmospheric mass is larger for a given surface temperature. Furthermore, the increasing rate of precipitation with increasing surface temperature under a larger atmospheric mass is smaller than that under a smaller atmospheric mass. These behaviors can be understood based on atmospheric or surface energy balance. Atmospheric mass influences Rayleigh scattering, multiple scattering in the atmosphere, pressure broadening, lapse rate, and thereby precipitation strength. These results have important implications on the climate and habitability of early Earth, early Mars, and exoplanets with oceans.
\end{abstract}

\section*{Plain Language Summary}
Precipitation is one of the key variables of the planetary climate system. Many factors can influence the strength of precipitation, such as solar flux, land-sea distribution, greenhouse gases, and aerosols. In this study, we show that another factor, atmospheric mass, can also strongly influence precipitation. The strength of precipitation increases with increasing surface temperature but decreases with increasing atmospheric mass. Furthermore, the increasing rate of precipitation with surface temperature becomes smaller under a larger atmospheric mass. These results have important implications for the climate evolution of early Earth, early Mars, and extra-solar rocky planets, which may have higher or lower air mass than that of modern Earth.

%% ------------------------------------------------------------------------ %%
%
%  TEXT
%
%% ------------------------------------------------------------------------ %%

\section{Introduction}
%\justifying
%Text here ===>>>
Precipitation is an important aspect of the climate system. Precipitation and its related latent heat release are one of the main driving factors of the atmospheric circulation on Earth \cite{holton2013introduction}. The spatial pattern of precipitation (as well as evaporation) and its variability influence the distribution of vegetation over the land \cite{lotsch2003coupled} and the surface salinity and density over the ocean \cite{huang1993real}. The seawater density further influences the global oceanic overturning circulation that transports energy from the tropical region to higher latitudes \cite{vallis2017atmospheric}. Furthermore, the strength of precipitation and its related runoff strongly affect the rate of carbonate-silicate weathering process, which determines atmospheric CO$_2$ concentration in geological timescales and then in turn influences the surface and air temperatures and precipitation \cite{walker1981negative,nagy2018dissolution}. The strength and spatial pattern of precipitation can be influenced by many factors, such as solar radiation, greenhouse gases, land-sea configuration, and aerosols. In this study, we emphasize another factor, atmospheric mass, which can strongly influence the strength of precipitation.

The atmospheric mass of Earth evolves with time. Before the first Great Oxidation Event  in $\sim$2.4 billion years ago (Gya), atmospheric O$_2$ concentration ($p$O$_2$) was negligible, and before the Neoproterozoic Oxidation Event in $\sim$0.6 Gya, it was about 10\% or less of the present level \cite{holland2006oxygenation}. For early Earth, \citeA{marty2013nitrogen}, \citeA{som2016earth}, and \citeA{avice2018evolution} showed that the surface pressure was likely less than the present level, and the lower limit is unconstrained but maybe smaller than 0.5 bar, but global nitrogen budget suggests that N$_2$ partial pressure might be $\ge$3 bar \cite{goldblatt2009nitrogen,johnson2015nitrogen,mallik2018oxidation}. For Mars, its surface pressure might be 1--2 bar in its early stage, but now it is only $\sim$0.006 bar \cite{lammer1991nonthermal}. For Venus, its surface pressure is close to 93 bar, mainly composed of CO$_2$ \cite{basilevsky2003surface}. Titan has a surface pressure of 1.45 bar \cite{niemann2005abundances}, but since the surface gravity of Titan is 1.35 m s$^{-2}$ ($\sim$0.14 of Earth's), its air mass per unit area is $\sim$10.5 times Earth's value. For planets beyond the solar system (or called exoplanets), their atmospheric masses, as well as compositions, can be quite different from those in the solar system \cite{turbet2018modeling,madhusudhan2019exoplanetary}.

Previous studies have shown that atmospheric mass has strong effects on planetary surface temperature \cite{goldblatt2009nitrogen,charnay2013exploring,le2014faint,kopparapu2014habitable,wolf2014controls,keles2018effect,xiong2020examining,zhang2020does,paradise2021climate} and on atmospheric circulation \cite<e.g.,>{kaspi2015atmospheric,chemke2016thermodynamic,chemke2017dynamics,guendelman2019atmospheric,komacek2019atmospheric,xiong2020possible}. Here, we focus on another climatic variable, surface precipitation. Using numerical simulations, \citeA{poulsen2015long} found that the strength of precipitation increases when $p$O$_2$ is reduced. They proposed that the mechanism is related to the reduction of Rayleigh scattering of air molecules under a lower $p$O$_2$ and thereby the increase of surface shortwave radiation. In this study, we extend their work by considering a much wider range of atmospheric mass. Furthermore, we explore the dependence of the increasing rate of precipitation with surface warming on atmospheric mass. For the mechanism, we find that atmospheric mass can influence not only shortwave radiation but also longwave radiation, both of which can influence the strength of precipitation, and the change of longwave radiation can significantly influence the increasing rate of global-mean precipitation with surface temperature.

%%%%%%%%%%%%%%%%%%%%%%%%%%%%%%%%%%%%%%%%%%%%%%%%%%%%%%%
%%%%%%%%%%%%%%%%%%%%%%%%%%%%%%%%%%%%%%%%%%%%%%%%%%%%%%%
%%%%%%%%%%%%%%%%%%%%%%%%%%%%%%%%%%%%%%%%%%%%%%%%%%%%%%%

\section{Model descriptions and experimental designs}

Our numerical simulations are mainly carried out using the Community Atmospheric Model version 3.0 \cite<CAM3, >{collins2004description,collins2006formulation} and the Exoplanet Community Atmospheric Model (ExoCAM). CAM3 is a widely used model developed by the National Center for Atmospheric Research (NCAR). The radiative transfer scheme in CAM3 is based on \citeA{ramanathan1986nonisothermal} and \citeA{briegleb1992delta}. The deep convection parameterization scheme is developed by \citeA{zhang1995sensitivity}. Clouds are treated as three types: marine stratus, convective clouds, and layered clouds. ExoCAM is developed based on CAM version 4.0 (CAM4). ExoCAM calculates radiative transfer with correlated-$k$ method based on the database of HITRAN~2004 \cite{wolf2013hospitable,wolf2015evolution,wolf2022exocam}. The CAM3 simulations have 26 vertical levels and the horizontal resolution is 2.8$^\circ$ $\times$ 2.8$^\circ$. In the ExoCAM simulations, we use 40 vertical levels with a horizontal resolution of 4$^\circ$ $\times$ 5$^\circ$. Eight main groups of experiments are designed, as listed in rows 2 to 9 of Table 1. For each group, we change the partial pressure of N$_2$ to test five different atmospheric masses, 0.5, 1, 5, 10, and 30 bar. In all the experiments, the gravity is 9.8 m\,s$^{-2}$.

In group A, we use CAM3 to perform the simulations, and surface temperature ($T_s$) is fixed with a meridional distribution of $T_s(\theta) = T_{ref} + \Delta T \times (1 - sin^{2}(\theta))$, where T$_{ref}$ is a reference temperature, which is varying from 275 K to 295 K with a 5-K interval, $\Delta T$ is the equator--pole temperature difference, and $\theta$ is latitude \cite{neale2000standard,medeiros2008aquaplanets}. Two different values of $\Delta T$ are tested, 20~and 50~K. The solar constant is 1365 W m$^{-2}$. The radiatively active gases include CO$_2$ (0.36 mbar), CH$_4$ (0.8 $\mu$bar), N$_2$O (0.27 $\mu$bar), and O$_3$ \cite<pre-industrial condition, see>{dutsch1978vertical}. The surface is ocean everywhere with no continent; this is called an aqua-planet.

%Table 1
 \begin{table}
 \caption{Models and Experimental Designs}
 \centering
 \small
 \begin{tabular}{lccccc}
 \hline\hline
   Groups & Model & Surface & Land--Sea & Surface Pressure & Descriptions\\
 \hline
  A &  CAM3 & Fixed-$T_s$ & Aqua-planet & 0.5, 1, 5, 10, 30 bar & Fixed S$_0$ and CO$_2$\\
  B &  CAM3 & Slab ocean & Aqua-planet & 0.5, 1, 5, 10, 30 bar & Fixed S$_0$, varied CO$_2$\\
  C &  CAM3 & Slab ocean & Aqua-planet & 0.5, 1, 5, 10, 30 bar & Varied S$_0$, fixed CO$_2$\\
  D &  CAM3 & Slab ocean & Aqua-planet & 0.5, 1, 5, 10, 30 bar & Varied S$_0$ and CO$_2$\\
  E &  CAM3 & Slab ocean & Modern continents & 0.5, 1, 5, 10, 30 bar & Fixed S$_0$, varied CO$_2$\\
  F &  CAM3 & Slab ocean & Modern continents & 0.5, 1, 5, 10, 30 bar & Varied S$_0$, fixed CO$_2$\\
  G &  CAM3 & Slab ocean & Modern continents & 0.5, 1, 5, 10, 30 bar & Varied S$_0$ and CO$_2$\\
  H &  ExoCAM & Fixed-$T_s$ & Aqua-planet & 0.5, 1, 5, 10, 30 bar & Fixed S$_0$ and CO$_2$\\
  \hline
 & ExoRT & Fixed-$T_s$ & Sea surface & 0.5, 1, 5, 10, 30 bar & 1-D radiative transfer model\\
 \hline
 Sens1 &  CAM3 & Fixed-$T_s$ & Aqua-planet & 10 bar & Weakened pressure broadening\\
 Sens2 &  CAM3 & Slab ocean & Aqua-planet & 10 bar & Change solar radiation\\
%Sens3 &  CAM3 & Fixed-$T_s$ & Aqua-planet & 1 and 5 bar & No CO$_2$, CH$_4$, or N$_2$O\\
%Sens4 &   ExoCAM & Slab ocean & Aqua-planet & 1 and 4 bar & Tide-locked planet\\
 Sens3 &   PlaSim & Slab ocean & Modern continents & 0.5, 1, 5, 10 bar & \citeA{paradise2021climate}\\
 Sens4 & SAM & Slab ocean & Sea surface & 0.5 and 1 bar & Small-domain, cloud-resolving\\
 \hline
 \end{tabular}
 \end{table}

In group B, the experiments are the same as in group A, except that the atmosphere is coupled to a slab ocean, rather than with fixed-$T_s$. The ocean depth is 50 m everywhere and no ocean heat transport is included. To obtain an appropriate temperature range, different CO$_2$ concentrations are performed for each surface pressure: for 0.5 bar, the CO$_2$ concentrations are 0.36, 0.72, 1.44, 2.88, and 5.76 mbar; for 1.0 bar, the CO$_2$ concentrations are 0.18, 0.36, 0.72, 1.44, and 2.88 mbar; for 5.0 bar, the CO$_2$ concentrations are 0.09, 0.18, 0.36, 0.72, and 1.44 mbar; and they are 0.045, 0.09, 0.18, 0.36, 0.72 mbar for 10.0 and 30.0 bar. The solar constant is fixed, 1365 W\,m$^{-2}$.

In group C, the experiments are similar to group B, but the partial pressure of CO$_2$ is fixed to 0.36 mbar, and the solar constant is varied: for 0.5 and 1.0 bar, the solar constant is 1330, 1365, 1400, 1430, and 1460 W\,m$^{-2}$; for 5.0 bar, the solar constants are 1290, 1320, 1340, 1365, and 1400 W\,m$^{-2}$; for 10.0 bar, the solar constants are 1270, 1300, 1330, 1365, and 1400 W\,m$^{-2}$; for 30.0 bar, the solar constants are 1250, 1275, 1300, 1330, and 1365 W\,m$^{-2}$.

In group D, the experimental design is similar to group B, but both CO$_2$ concentration and solar constant are varied. Five different CO$_2$ concentrations are performed: 0.18, 0.36, 0.72, 1.44, and 2.88 mbar, for all the five surface pressures. We have also changed the insolation to keep $T_s$ in an appropriate range. For the 0.5 bar surface pressure, the solar constant is 1400 W\,m$^{-2}$, and 1365 W\,m$^{-2}$ for 1 bar, 1320 W\,m$^{-2}$ for 5 bar, 1300 W\,m$^{-2}$ for 10 bar, and 1250 W\,m$^{-2}$ for 30 bar.

Group E is the same as group B, group F is the same as group C, and group G is the same as group D, except that modern Earth’s land-sea distribution is used, rather than an aqua-planet mode. These three groups are performed to test the effect of land--sea configuration.

In group H, the experiments are the same as group A but use the model ExoCAM. Comparing group H with group A, we can know whether our conclusion is robust under different radiative transfer schemes. Briefly, the greenhouse effect of water vapor in ExoCAM is stronger, and shortwave absorption by water vapor in ExoCAM is also stronger than in CAM3. For more details, please see \citeA{yang2016differences}, \citeA{kumar2017habitable}, and \citeA{wolf2022exocam}.

To clarify the underlying mechanism, we have also done a series of 1-D radiative transfer calculations with the Exoplanet radiative transfer model \cite<ExoRT,>{wolf2017constraints} and several groups of GCM sensitivity experiments (see Table~1). Using ExoRT, five different surface pressures are considered: 0.5, 1, 5, 10, and 30 bar. Under each surface pressure, $T_s$ is varied from 285 to 305 K with 5-K intervals. The temperature structures are moist adiabatic profiles, with a 150-K isothermal stratosphere. In the troposphere, the atmosphere is set to be saturated. In the stratosphere, the specific humidity is set to be equal to the value at the top of the troposphere. For the details and results of the GCM sensitivity experiments, please see section~3.3 below.

In the GCM experiments of fixed-$T_s$, the model takes less than 10 model years to reach equilibrium. We run the experiments for 15 years. Data of the last 5 model years are used to analyze below. In the GCM experiments of slab ocean, the model takes about 30 model years to reach equilibrium. Each experiment was run for at least 40 years, and the last 10 years are analyzed. In all the 3-D simulations, the vertical profiles of temperature, water vapor, clouds, and other variables are actively coupled to radiation, advection, and convection processes, regardless if the fixed-$T_s$ or slab ocean setup was used. After reaching equilibrium, annual-mean net atmospheric energy imbalance is within $\pm$0.5 W~m$^{-2}$ in all the experiments. Meanwhile, time series of surface temperature and precipitation fields were analyzed and they have also reached equilibrium.

%%%%%%%%%%%%%%%%%%%%%%%%%%%%%%%%%%%%%%%%%%%%%%%%%%%%%%%
%%%%%%%%%%%%%%%%%%%%%%%%%%%%%%%%%%%%%%%%%%%%%%%%%%%%%%%
%%%%%%%%%%%%%%%%%%%%%%%%%%%%%%%%%%%%%%%%%%%%%%%%%%%%%%%

\section{Results}
\subsection{Dependence of precipitation-$T_s$ relationship on atmospheric mass}

% Figure 1
\begin{figure}[htbp]
\noindent\includegraphics[width=\textwidth]{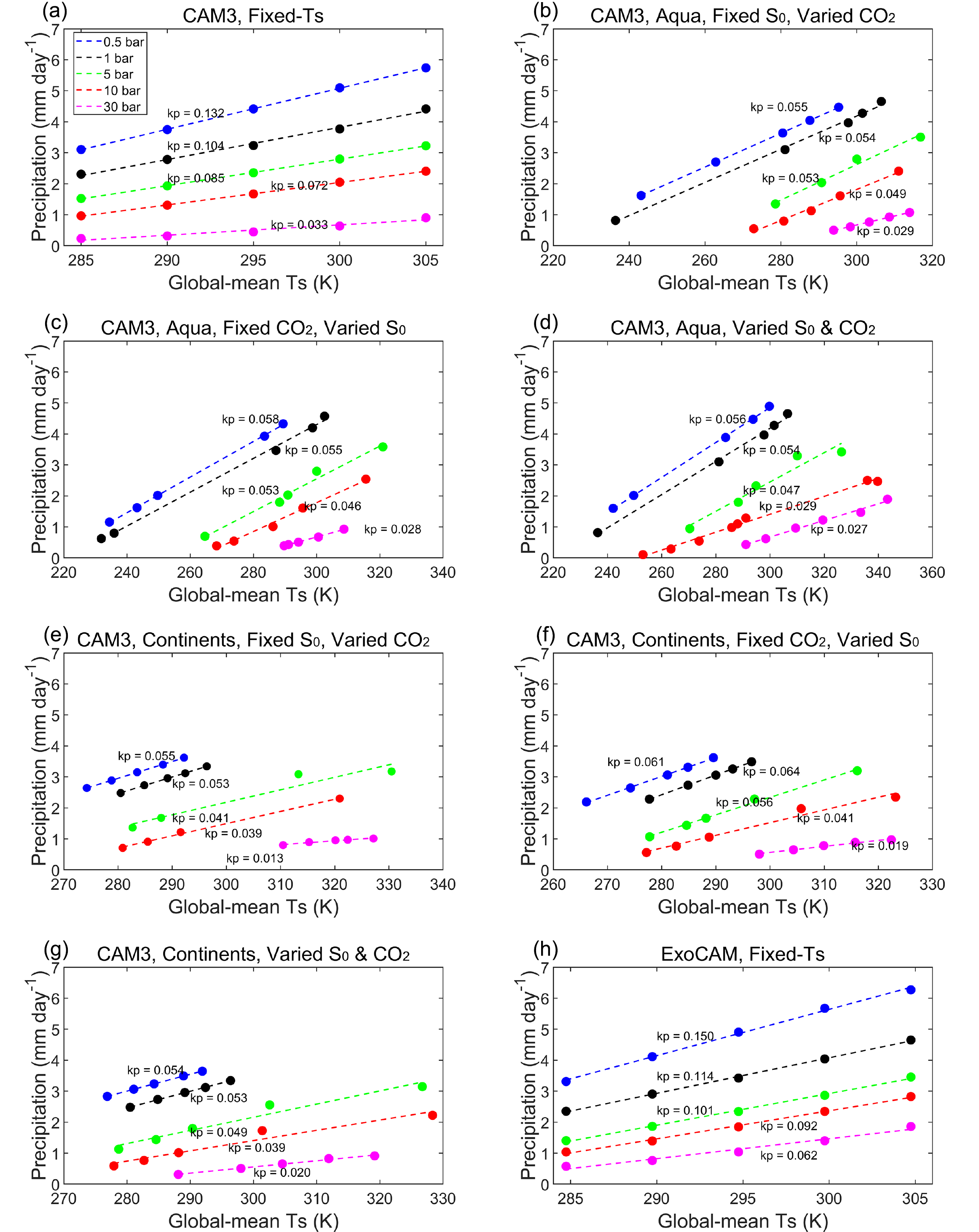}
\caption{Annual- and global-mean precipitation as a function of global-mean $T_s$ in the eight groups of experiments. (a) group A in which $T_s$ is fixed, the model is CAM3, and it is in an aqua-planet mode (see also Figure~S9). (b) group B, same as (a) but the surface is actively coupled to a 50 m slab ocean, CO$_2$ concentration is varied but the solar constant is fixed (1365 W\,m$^{-2}$). (c) group C, same as (b) but CO$_2$ concentration is fixed to 0.36~mbar and the solar constant is varied. (d) group D, same as (b) but both CO$_2$ concentration and solar constant are varied. (e)-(g) are same as (b)-(d) but with modern Earth's land-sea distribution. (h) group H, same as (a) but using the model ExoCAM. The dashed lines are the linear regressions. The slope of each regression line is listed near the line with units of mm\,day$^{-1}$\,K$^{-1}$. Two main conclusions: 1) the precipitation is weaker under a higher air mass for a given $T_s$, and 2) the precipitation-$T_s$ slope is flatter under a higher air mass, although the absolute values differ significantly among the different groups.}
\label{precip}
\end{figure}

Figure~\ref{precip} shows annual- and global-mean precipitation as a function of global-mean $T_s$ in all eight groups of experiments. Due to different surface boundary conditions (fixed-$T_s$ or coupled to a slab ocean, and with or without continents), different surface temperature ranges, different external forcings (varied stellar radiation, CO$_2$ concentration, or both), and different models (CAM3 and ExoCAM), the absolute strengths of precipitation (including both rainfall and snowfall) are different between the groups. But, all these eight groups show three clear trends: 1) the precipitation is a nearly linear increasing function of $T_s$; 2) under a given global-mean $T_s$, the precipitation is weaker when the surface pressure is higher; 3) the slope of precipitation--$T_s$ strongly depends on the value of surface pressure, and it decreases under a higher surface pressure. The first trend is not new and it has been found in many previous studies of global warming such as \citeA{trenberth2005relationships}, \citeA{lambert2008dependency}, and \citeA{yang2016monotonic}. So, in the following, we will focus on the second and third trends.

Figures S1--S8 show the spatial patterns of annual-mean precipitation in the eight groups of experiments. There are two precipitation-concentrated regions, one in the deep tropics where there is strong intertropical convergence and one in middle latitudes where there is strong baroclinic instability. The three trends listed above can be found in both these two regions. This implies that the trends do not depend on the detailed dynamical processes (convergence, instability, monsoon, or storm track) or the type of convection (deep, shallow, or mixed). There should be some unified mechanism that determines the behavior of the mean precipitation.

\subsection{Mechanism}

\begin{figure}[htbp]
\centering
\noindent\includegraphics[width=\textwidth]{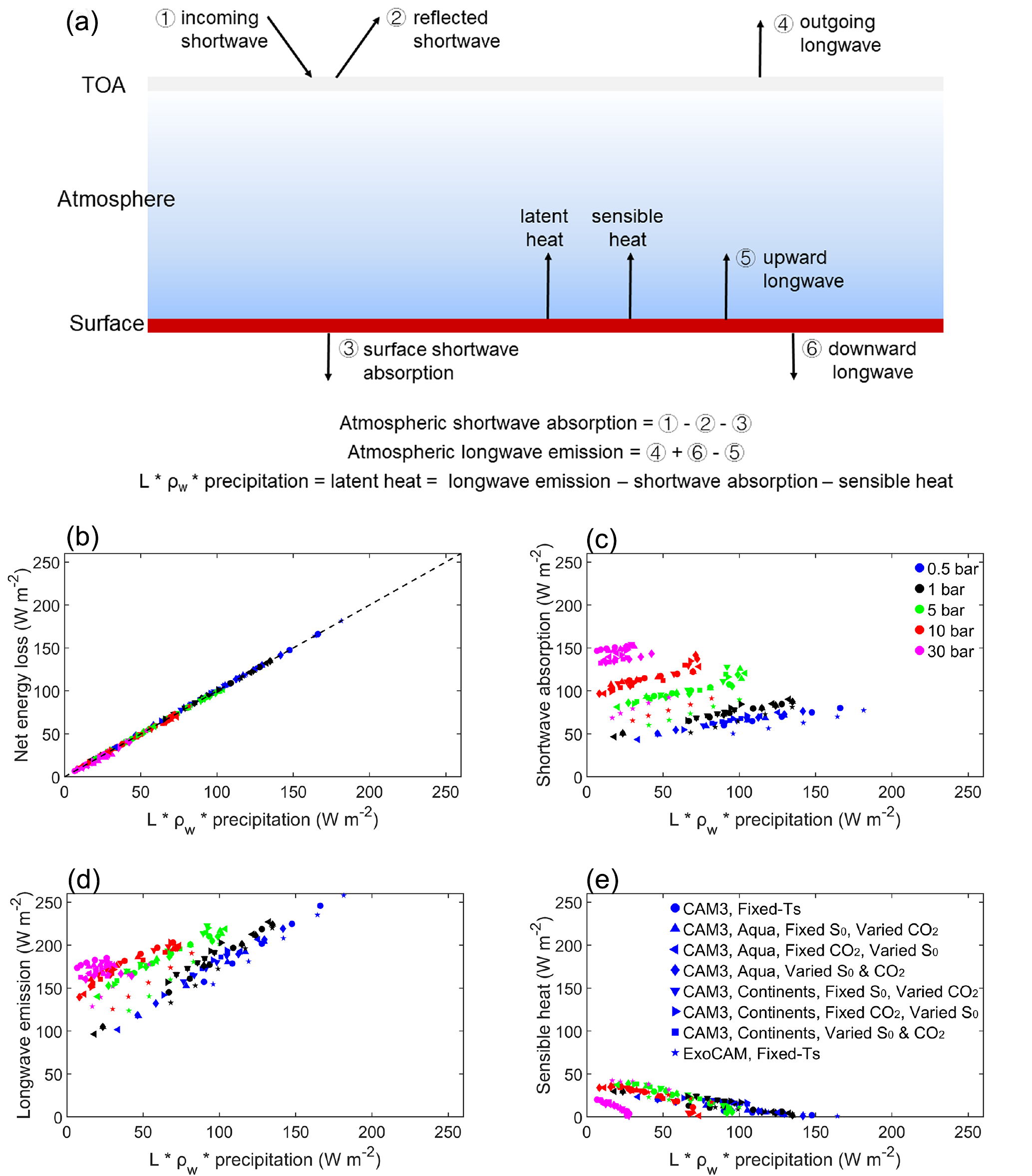}
\caption{(a) Schematic diagram of how global-mean precipitation is determined by energy budget in the system. In global-mean, latent heat release (associated with precipitation) is equal to atmospheric net energy loss, which is equal to the combination of atmospheric longwave emission, atmospheric shortwave absorption, and surface sensible heat flux. (b) Atmospheric net energy loss vs precipitation; (c) atmospheric shortwave absorption vs precipitation; (d) atmospheric net longwave emission vs precipitation; and (e) surface sensible heat flux vs precipitation. The global-mean precipitation has been multiplied by $L$ $\rho_w$, where $L$ is the specific heat of water vaporization, and $\rho_w$ is the density of liquid water. The definition of the directions of the different variables in panels (b)-(e) follows panel (a). This figure shows that the precipitation strength is constrained by net energy loss in the atmosphere, regardless of the surface type, fixed-$T_s$ or coupled to a slab ocean.}
\label{schematic}
\end{figure}

In the analyses of precipitation change under global warming, many studies have found that atmospheric or surface energy balance is a good approach, which can quantitatively estimate the change of precipitation under different conditions \cite<e.g.,>{pierrehumbert2002hydrologic,allen2002constraints,o2012energetic,showman2013atmospheric,pendergrass2014atmospheric,jeevanjee2018mean}. A diagram of the atmospheric energy budget is shown in Figure~\ref{schematic}a. The air receives shortwave radiation from the star, reflects a part of the shortwave radiation back to space, obtains longwave radiation, sensible heat, and latent heat from the surface, and meanwhile emits longwave radiation to space and surface. In the long-term and global mean, the amount of latent heat that acts to heat the atmosphere is equal to the amount of atmospheric net energy loss (i.e., atmospheric net longwave emission $-$ atmospheric shortwave absorption $-$ sensible heat). Meanwhile, the value of net latent heat release is equal to the amount of surface precipitation ($P$) multiplied by the specific heat of water vaporization ($L$) and the density of liquid water ($\rho_w$), written as:
\begin{linenomath*}
\begin{equation}
L\rho_{w}P = LH = LW^a - SW^a - SH
\end{equation}
\end{linenomath*}
where $LH$ is surface latent heat flux, $LW^a$ is atmospheric net longwave emission, $SW^a$ is atmospheric shortwave absorption, and $SH$ is surface sensible heat flux. All these terms are global-mean values, so horizontal heat and water transports are not explicitly included.

%Figure 2

Figure~\ref{schematic}b shows the relationship between the atmospheric net energy loss and the precipitation rate multiplied by $L$ and $\rho_{w}$. All the experiments of the eight groups fall in the 1-to-1 diagonal line; this confirms that the above atmospheric energy budget is correct and also implies that the atmosphere has already reached equilibrium in all the experiments. In the fixed-$T_s$ experiments, the surface is not energy-balanced, but the net energy gain or loss at the surface is equal to that at the top of the atmosphere (TOA, figures not shown). Figures~\ref{schematic}c, \ref{schematic}d,~\&~\ref{schematic}e show the atmospheric shortwave absorption, atmospheric net longwave emission, and surface sensible heat flux, respectively. It can be seen that the atmosphere net longwave emission dominates, and the sensible heat flux is the smallest but not negligible. This is similar to that on modern Earth, for which the three values are $\approx$ 80, 188, and 20 W\,m$^{-2}$, respectively \cite{hartmann2015global}. In the CAM3 experiment with modern continents, CO$_2$ concentration, and solar flux (in the group F), the global-mean surface temperature is 289~K and the three values are 74, 180, and 20 W\,m$^{-2}$, respectively.

\begin{figure}[htb]
\noindent\includegraphics[width=\textwidth]{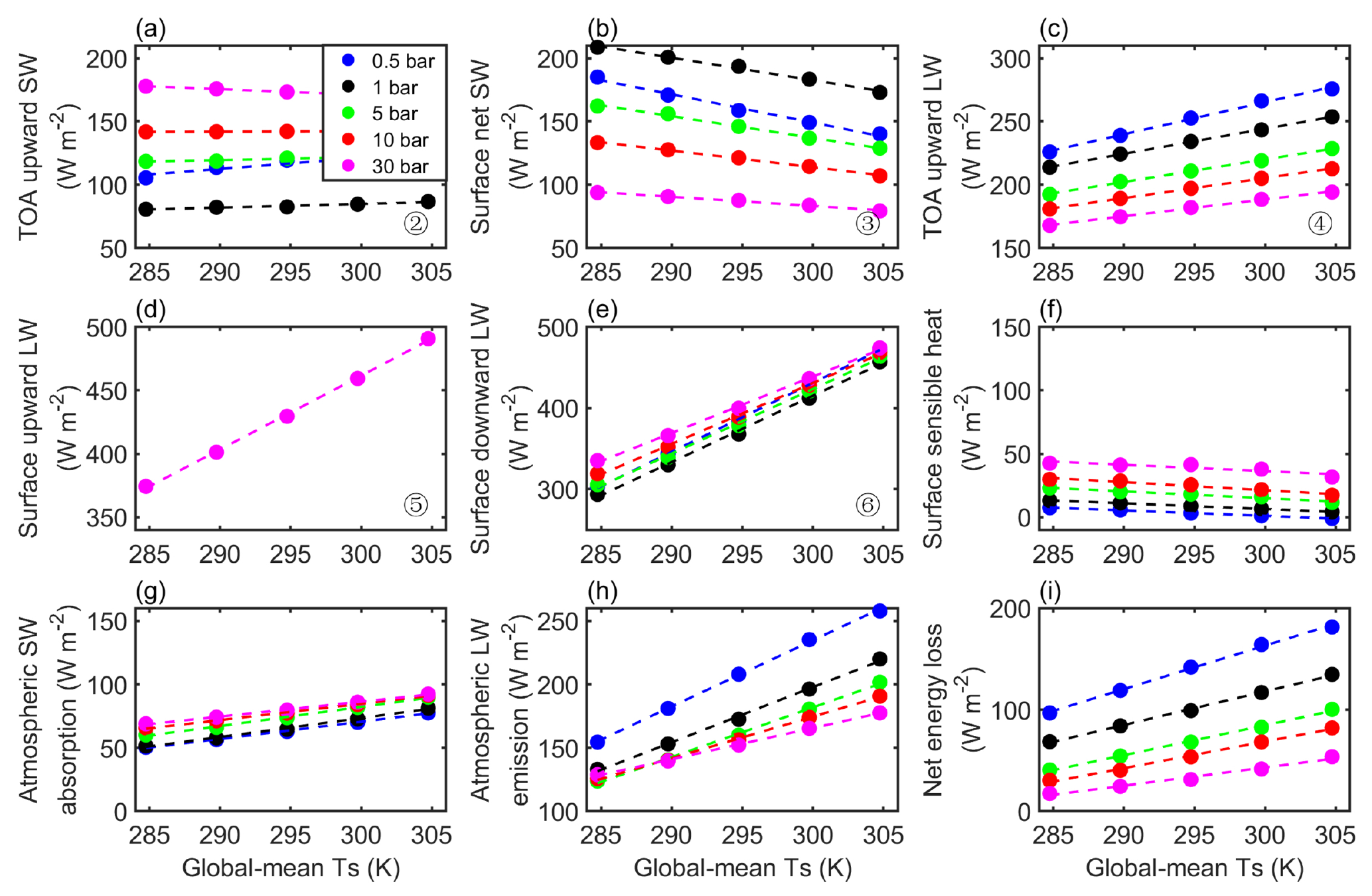}
\caption{Global-mean atmospheric energy budget as a function of $T_s$ in the group H experiments. (a) Upward shortwave (SW) flux at the top of the atmosphere (TOA), (b) net shortwave flux at the surface, (c) upward longwave (LW) flux at the TOA, (d) upward longwave flux at the surface, (e) downward longwave flux at the surface,  (f) surface sensible heat flux, (g) atmospheric shortwave absorption, (h) atmospheric longwave emission, and (i) atmospheric net energy loss. The definition of the directions of the different variables follows Figure~\ref{schematic}a. In panel (d), the values for different surface pressures are the same because the surface temperatures are the same (specified). The numbers of \ding{193} to \ding{197} in panels (a)--(e) are corresponding to those in Figure~\ref{schematic}a. Precipitation strength is constrained by atmospheric net energy loss (see equation (1)).}
\label{flux}
\end{figure}

Why does the precipitation rate become smaller under a larger atmospheric mass even under the same global-mean $T_s$? The answer is shown in Figure~\ref{flux}. There are two reasons. One is that the atmospheric net longwave emission decreases and the other one is that the atmospheric shortwave absorption mainly by water vapor in near-infrared wavelengths increases, as air mass is increased.

The former, the change of the atmospheric net longwave emission, is because the atmospheric greenhouse effect becomes stronger with increasing air pressure, which in turn, is mainly because of the effect of pressure broadening and the change of vertical temperature profile. As the surface pressure increases under a given surface temperature, upward longwave emission to space decreases (Figures~\ref{flux}c~\&~S10), downward longwave radiation to the surface increases (Figure~\ref{flux}e), and the change of the former is larger than that of the latter, so that the atmospheric net longwave emission decreases (Figure~\ref{flux}h). These trends can also be found in 1-D off-line radiative transfer calculations, as shown in Figures~S11, S12~\&~S13. In longwave wavelengths, the effect of increased pressure broadening is qualitatively equivalent to adding more CO$_2$ to the atmosphere. Both of them act to increase the atmospheric greenhouse effect by trapping more thermal radiation from the surface and then re-emitting back to the surface, i.e., the thermal transmissivity of the atmosphere decreases \cite<see Text~S1, Figure~S14 in Supporting Information, and>{li1997atmospheric}. Meanwhile, a higher air mass increases the atmospheric heat capacity and thereby makes the lapse rate larger (i.e., closer to dry adiabat, Figure~S11, S15~\&S16, see also such as \citeA{nakajima1992study}, \citeA{goldblatt2009nitrogen} and \citeA{xiong2020possible}). This leads to a stronger atmospheric greenhouse effect and less atmospheric longwave emission to space, which leads to weaker precipitation according to equation (1).

The latter, the increase of atmospheric shortwave absorption, is because increasing air mass enhances multiple scattering in the atmosphere, which means that the shortwave absorption path of water vapor becomes longer (Figures~\ref{flux}g \& \ref{schematic}c). Meanwhile, when air mass is increased, more shortwave energy is reflected back to space due to enhanced Rayleigh scattering (Figures~\ref{flux}a \& S17), so less shortwave energy reaches the surface (Figure~\ref{flux}b); based on this, \citeA{poulsen2015long} suggested that surface evaporation (and precipitation) should become smaller, because the global-mean evaporation should be equal to surface shortwave absorption minus net surface longwave emission and surface sensible heat flux (the framework of surface energy budget). These results are consistent with off-line 1-D radiative-transfer calculations (Figure~S17, see also \citeA{lacis1991description}, \citeA{paynter2011assessment}, and \citeA{goldblatt2016comment}). Moreover, from Figures~\ref{flux}a,~\ref{flux}b,~\&~\ref{flux}g, one could find that in these experiments the changes of the shortwave flux at the TOA and at the surface are about 5--10 times that of the changes of the absorbed shortwave flux in the atmosphere; the former is the component of TOA or surface energy budget, the latter is the component of atmospheric energy budget, and there is no contradiction between them.

% Figure 3

Why does the precipitation--$T_s$ slope become smaller under a larger atmospheric mass? The answer is also shown in Figure~\ref{flux}, in particular, the upward longwave emission at the TOA or called outgoing longwave radiation (OLR) shown in Figure~\ref{flux}c. OLR is contributed from two parts: one is the portion of the emission from the surface which is transmitted by the entire atmosphere, and the other one is from the atmosphere itself \cite{pierrehumbert2010principles}. OLR is a nearly linear increasing function of $T_s$ over a wide range of temperatures, unless the surface is too cold (such as below 190~K) or too hot (such as above 320~K). This is because of the competition between the increased surface thermal emission with temperature and the narrowing of spectral window regions ($\approx$\,8--12~$\mu$m) related to the increase in atmospheric water vapor concentration \cite{koll2018earth,zhang2020linearity}. The slope of the OLR--$T_s$ relation depends on the surface temperature and the average transmissivity of the atmosphere \cite<equation~[3] and Figure~4 in>{koll2018earth}. In our experiments, the transmissivity is lower under a higher atmospheric mass (Figure~S14), so the OLR--$T_s$ slope becomes smaller. Another reason for the decrease of the OLR--$T_s$ slope is that the increase of air temperature of the upper troposphere under a higher surface pressure is smaller than that under a lower surface pressure, for the same $T_s$ change (as shown in Figure~S11, S15~\&~S16). This also makes the sensitivity of OLR to $T_s$ become smaller (Figure~\ref{flux}c~\&~S10). Again, the underlying mechanism relates to that the temperature profile is steeper under a larger air mass (see Text~S2 in Supporting Information).

The decreasing trend of the OLR--$T_s$ slope with increasing atmospheric mass was also found in previous clear-sky radiative transfer calculations, such as shown in Figure~5b of \citeA{goldblatt2013low}, Figure~1a of \citeA{kopparapu2014habitable}, and Figure~1a in \citeA{zhang2020does}. Note that this trend occurs only when the atmosphere is below the runaway greenhouse limit. In the runaway greenhouse state, the atmosphere becomes optically thick in all thermal wavelengths due to the continuum absorption of water vapor and thereby OLR nearly does not depend on $T_s$ \cite{koll2018earth}. In other words, for very optically-thick conditions, the upper troposphere becomes the primary emission layer and subsequently OLR is decoupled from the surface.

Figures S13a~\&~S18 show that for a given surface temperature, the amount of vertically-integrated water vapor is lower under a higher air mass (because the atmospheric temperature profile is closer to dry adiabat), but the corresponding value of OLR is smaller rather than greater in both 3-D GCMs and 1-D radiative transfer models (Figures~\ref{flux}c,~S10,~\&~S13c). This is counter-intuitive as it is well known that water vapor has a strong greenhouse effect and in general a lower water vapor content should be corresponding to a greater OLR. But, the pressure broadening effect of the changed air mass is strong and the lapse rate is larger (closer to dry adiabat) for a higher air mass, which acts to trap more thermal radiation and decrease the value of OLR.

The third term on the right side of equation (1) is surface sensible heat flux. In the model, sensible heat flux is calculated by $\rho$$c_p$$C_D$$W$$(\theta_s-\theta_a)$, where $\rho$ is the near-surface air density, $c_p$ is the specific heat capacity of air, $C_D$ is stability-dependent turbulent exchange coefficient, $W$ is surface wind speed, and $\theta_s$ and $\theta_a$ are the potential temperatures of surface and near-surface air, respectively. Surface wind speed decreases with increased atmospheric mass, due to the decreases of horizontal temperature gradients under massive atmospheres, which in turn is related to the increase of meridional energy transport under a higher air mass (Figure S19; for more details, please see \citeA{chemke2016thermodynamic} and \citeA{chemke2017dynamics}). Meanwhile, the temperature difference $(\theta_s$\,$-$\,$\theta_a)$ also decreases under a higher air mass. The near-surface air density, however, increases significantly (figures not shown). As a result, higher air mass leads to larger sensible heat flux, as shown in Figure~\ref{flux}f. This also implies that the precipitation should become weaker under a higher atmospheric mass, based on the energy balance constraint in equation (1). But, the change of the sensible heat--$T_s$ slope under different atmospheric masses is very small (Figure~\ref{flux}f), so that the precipitation--$T_s$ slope is not influenced by the changes of the sensible heat flux.

%%%%%%%%%%%%%%%%%%%%%%%%%%%%%%%%%%%%%%%%%%%%%%%%%%
%%%%%%%%%%%%%%%%%%%%%%%%%%%%%%%%%%%%%%%%%%%%%%%%%%
%%%%%%%%%%%%%%%%%%%%%%%%%%%%%%%%%%%%%%%%%%%%%%%%%%

\subsection{Sensitivity experiments and results}

\begin{figure}[htbp]
\noindent\includegraphics[width=\textwidth]{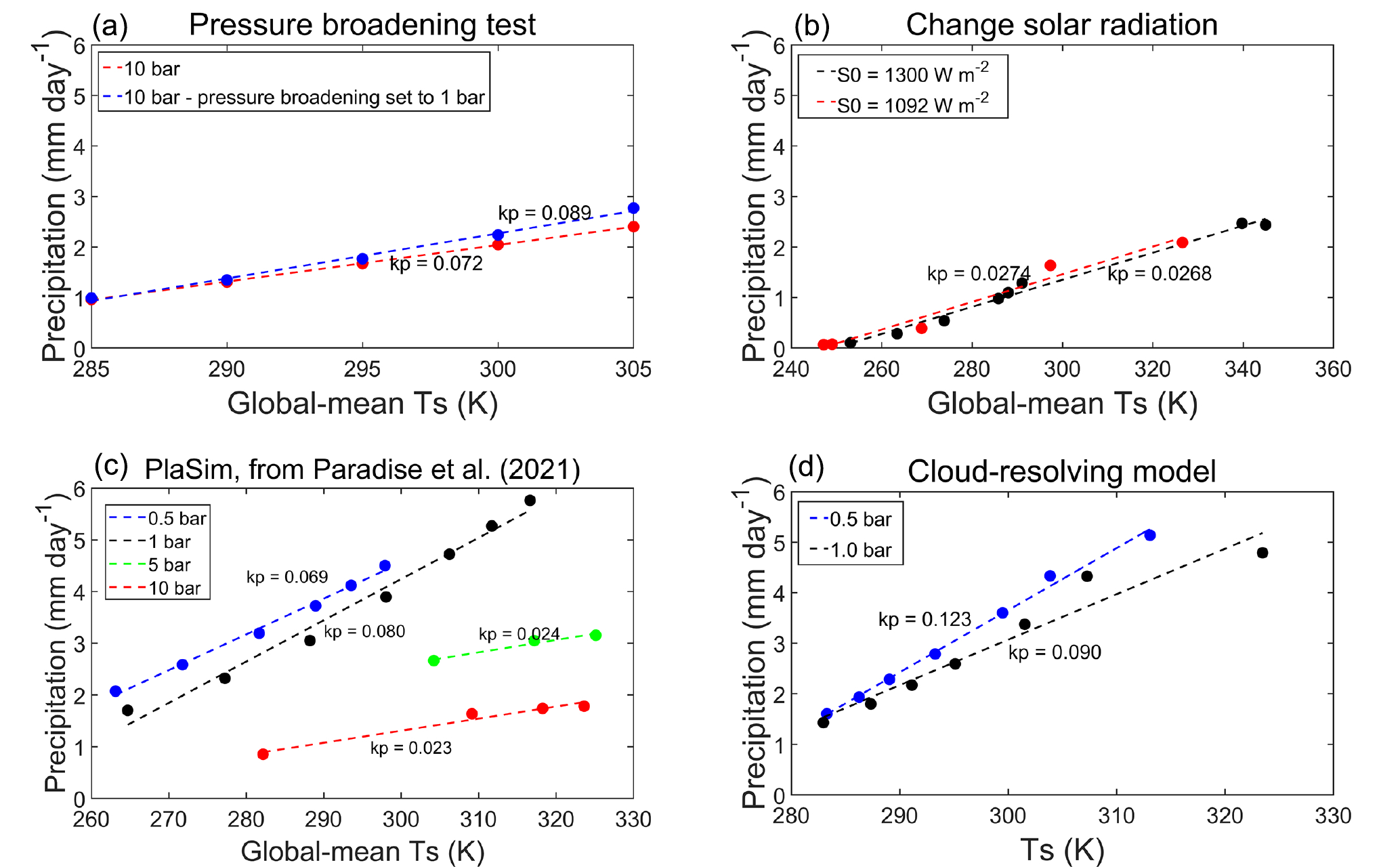}
\caption{Results of the sensitivity experiments. (a) The run of 10.0 bar surface pressure (red line) and the sensitivity run of 10.0 bar with pressure broadening set to 1 bar (blue line), using CAM3 with fixed-$T_s$; (b) the runs of changing solar constant from 1300 to 1092 W\,m$^{-2}$, using CAM3 coupled to a 50-m slab ocean; (c) the results of another GCM PlaSim with data from \citeA{paradise2021climate}, and (d) the results of cloud-resolving simulations without convection parameterization, using SAM coupled to a slab ocean. The lines are for linear regressions. These sensitivity experiments confirm the two conclusions found in Figure~1.}
\label{sensitivitytest}
\end{figure}

In order to further confirm the mechanisms, we perform four sensitivity groups of experiments, as listed in rows 11 to 14 of Table 1. As addressed in section 3.2, pressure broadening is one of the important factors, which influence longwave radiative transfer and thereby the precipitation strength. To more clearly examine this, we add one group of experiments within which the surface pressure is 10~bar but the effect of pressure broadening is set to that of 1 bar. When the pressure broadening is weakened, the slope of the precipitation--$T_s$ relationship increases from 0.072 to 0.089 mm\,day$^{-1}$\,K$^{-1}$ (Figure~\ref{sensitivitytest}a). This confirms that pressure broadening has an effect of reducing the precipitation--$T_s$ slope.

Also addressed in section 3.2, shortwave radiation can influence the strength of precipitation but nearly does not influence the precipitation--$T_s$ slope. To further confirm this, we add one group of experiments within which the solar constant is changed from 1300 to 1092 W\,m$^{-2}$. The precipitation--$T_s$ slope almost stays unchanged (Figure~\ref{sensitivitytest}b). This confirms that shortwave radiation is not the key factor that influences the precipitation--$T_s$ slope but it does influence $T_s$ and the absolute value of precipitation.

% Figure 4

Besides CAM3 and ExoCAM, we find that our main results can also be verified in another GCM, PlaSim. The results of PlaSim are shown in Figure~\ref{sensitivitytest}c, and the data was downloaded from the Borealis repository \cite{paradise2021climate}. \citeA{paradise2021climate} employed the model to examine the effects of varying N$_2$ partial pressure on surface temperature, sea ice coverage, clouds, and atmospheric circulation. PlaSim is coupled to a 50-m slab ocean; the convection scheme and radiative transfer scheme of PlaSim are relatively simpler than CAM3 and ExoCAM; the model resolution, 5.625$^{\circ}$ in latitude and 5.625$^{\circ}$ in longitude, is lower than CAM3 and ExoCAM; and modern Earth’s land-ocean configuration was used in their simulations. Although the radiative transfer module, convection parameterization, and cloud and precipitation schemes in PlaSim are different from CAM3 or ExoCAM, the basic effects of varying air mass on Rayleigh scattering, pressure broadening, and lapse rate are included. So, it is not a surprise to view the same conclusions in PlaSim.

In CAM3, ExoCAM, and PlaSim, convection and clouds are parameterized. The model resolutions are not fine enough to resolve convection processes. In the final sensitivity group, we use a cloud-resolving model, the System of Atmospheric Modeling \cite<SAM,>{khairoutdinov2003cloud}. SAM uses a non-hydrostatic dynamic core and employs the radiative transfer module of RRTM \cite<a rapid and accurate radiative transfer model,>{mlawer1997radiative} that is similar to that used in ExoCAM, but ExoCAM can more accurately simulate extremely hot and wet climates such as runaway greenhouse \cite{wolf2022exocam}. We employ a domain of 96 km $\times$ 96 km, and the horizontal resolution is 1 km $\times$ 1 km. The time step is 10 s. Two surface pressures are tested, 0.5 and 1 bar. Each SAM experiment runs for 2500 model days. The outputs of the last 100 days are analyzed. Under 0.5 bar, the precipitation--$T_s$ slope is 0.123 mm\,day$^{-1}$\,K$^{-1}$, but it decreases to 0.090 mm\,day$^{-1}$\,K$^{-1}$ when surface pressure is 1 bar (Figure~\ref{sensitivitytest}d). These results indicate that our conclusions do not rely on the convection/cloud schemes.

%%%%%%%%%%%%%%%%%%%%%%%%%%%%%%%%%%%%%%%%%%%%%%%%%%%%%%%
%%%%%%%%%%%%%%%%%%%%%%%%%%%%%%%%%%%%%%%%%%%%%%%%%%%%%%%
%%%%%%%%%%%%%%%%%%%%%%%%%%%%%%%%%%%%%%%%%%%%%%%%%%%%%%%

\section{Conclusions and Discussions}

In this study, we have explored the dependence of global-mean precipitation on atmospheric mass and its sensitivity to $T_s$ change. First, the strength of precipitation decreases with air mass under a given $T_s$. Second, the slope of the precipitation--$T_s$ relationship becomes flatter under a higher air mass. These trends can be understood based on atmospheric or surface energy balance budget. The first result is contributed by the combined effect of increased Rayleigh scattering, enhanced atmospheric shortwave absorption, weakened net atmospheric longwave emission, and increased surface sensible heat flux as air mass is larger. The latter is due to that longwave atmospheric transmissivity becomes smaller and meanwhile the lapse rate becomes steeper when air mass is increased.

This study emphasizes a new factor, atmospheric mass, which can influence the sensitivity of global-mean precipitation to $T_s$. Over geological time scales or on other rocky planets, the atmospheric masses can be significantly different from modern Earth. Different atmospheric masses influence the response of precipitation to $T_s$ and therefore the timescale of the carbonate--silicate weathering cycle. This coupled process can further influence global climate, its evolution, and planetary habitability. In particular, these results suggest that precipitation is very weak on planets having massive atmospheres, which can strongly limit planetary habitability.

In this research, we explain the response of precipitation to air mass through energy balance budget. But, atmospheric dynamics may offer another way to understand the results. For example, when atmospheric mass increases, the strengths of the meridional circulation and convection would decrease \cite{chemke2016thermodynamic,chemke2017dynamics}. The extra-tropical eddies also become weaker because of less mean eddy kinetic energy per unit mass \cite{chemke2017dynamics}. The advantage of the approach we used here is that energy balance budget is  more quantitative and easier to analyze.

Figures 1 and 4 show that there are large quantitative differences in the precipitation strength and the precipitation--$T_s$ slope between aqua-planets and planets with modern Earth’s continents. Also, there are significantly quantitative differences between the three models, CAM3, ExoCAM, and PlaSim. It is important to know what exact mechanisms/processes cause the differences in the future. These results also suggest that in studying precipitation under different climate states, one model is not enough and a hierarchy of models or multiple models are required.

In this study, we focus on the variable of precipitation. But, other related variables of the climate system, such as cloud fraction, cloud water path, and water vapor concentration (Figure~S18), are also influenced by atmospheric mass. These variables can strongly influence observational atmospheric characteristics such as transmission, emission, and reflection spectra. Moreover, in our experiments, the land water storage has also changed, which can influence the types of surface vegetation if it existed on exoplanets. All these are left for future work.

More experiments in the future are required to examine whether the conclusions are applicable to other planets, such as planets with different background gases (such as H$_2$, He, O$_2$, CO, or CO$_2$), planets having different gravities, different stellar spectra, and tidally locked planets with different rotation rates or spin-orbit resonances. In our experiments, air mass, the number of molecules, and surface air pressure are changed, but both planetary gravity (9.8 m\,s$^{-2}$) and background atmospheric molecular weight (28 g mole$^{-1}$, N$_2$) are fixed. The separated effects of these different factors are not explicitly addressed in this article. For example, Rayleigh and Mie scatterings depend on the number of molecules and the inherent properties of molecules (such as N$_2$ versus CO$_2$), the strength of pressure broadening depends on air mass and planetary gravity, and the molecular weight compared to H$_2$O can strongly influence the strength of buoyancy for convection. Previous studies had examined the effect of varying planetary gravity on radiation transfer, surface temperature, and atmospheric dynamics \cite<such as>{pierrehumbert2010principles,kaspi2015atmospheric,kilic2017impact,yang2019effects,yang2019planetary,thomson2019effects}, but no work has explored its effect on surface precipitation. Moreover, another important process related to precipitation strength is shortwave absorption, which is strongly influenced by water vapor concentration, clouds, and stellar spectrum. Water vapor and clouds have greater absorption coefficients in near-infrared wavelengths. If the star has a redder spectrum, more stellar energy would be absorbed by the atmosphere (rather than by the surface), therefore the atmosphere would be more stable \cite<e.g.,>{eager2020implications} and convection and precipitation would be weaker. All these should be further examined in future studies.

%%%%%%%%%%%%%%%%%%%%%%%%%%%%%%%%%%%%%%%%%%%%%%%%%%%%%%%
%%%%%%%%%%%%%%%%%%%%%%%%%%%%%%%%%%%%%%%%%%%%%%%%%%%%%%%
%%%%%%%%%%%%%%%%%%%%%%%%%%%%%%%%%%%%%%%%%%%%%%%%%%%%%%%

\section*{Open Research}
The simulation data is available at: \url{https://doi.org/10.5281/zenodo.7011993}. The CAM3 code is available at: \url{www.cesm.ucar.edu/models/atm-cam/}. The CESM code is available at: \url{www.cesm.ucar.edu/models/cesm1.2/}. The ExoCAM and ExoRT are available at: \url{https://github.com/storyofthewolf/ExoCAM}. The SAM code is available at: \url{http://rossby.msrc.sunysb.edu/~marat/SAM.html}. The PlaSim simulation data can download from from the Borealis repository: \url{https://doi.org/10.5683/SP2/LOFVCV} \cite{paradise2021climate}.

%%%%%%%%%%%%%%%%%%%%%%%%%%%%%%%%%%%%%%%%%%%%%%%

\acknowledgments
Jun Yang is supported by the National Natural Science Foundation of China (NSFC) under grants 42075046 and 42161144011. Our simulations use about 3.0$\times$10$^5$ core-hours in total, and this is corresponding to CO$_2$ emission of about 4.8$\times$10$^2$ kg, if assume the power per core is 3.7 W and the average carbon emission intensity is around 0.7 kg per kWh.

%% ------------------------------------------------------------------------ %%
%% References and Citations

%%%%%%%%%%%%%%%%%%%%%%%%%%%%%%%%%%%%%%%%%%%%%%%
%
% \bibliography{<name of your .bib file>} don't specify the file extension
%
% don't specify bibliographystyle

% In the References section, cite the data/software described in the Availability Statement (this includes primary and processed data used for your research). For details on data/software citation as well as examples, see the Data & Software Citation section of the Data & Software for Authors guidance
% https://www.agu.org/Publish-with-AGU/Publish/Author-Resources/Data-and-Software-for-Authors#citation

%%%%%%%%%%%%%%%%%%%%%%%%%%%%%%%%%%%%%%%%%%%%%%%

\bibliography{refs}

\end{document}